\documentclass[%
 reprint,
superscriptaddress,
 amsmath,amssymb,
 aps,
 showkeys
]{revtex4-2}

\usepackage{graphicx}
\usepackage{dcolumn}
\usepackage{bm}
\usepackage[pdfborder={0 0 0}]{hyperref}

\usepackage{physics}
\usepackage{overpic}
\usepackage{capt-of}

\usepackage{xcolor}

\usepackage{comment}

\begin{document}

\preprint{APS/123-QED}


\title{Quantum repeater segment with free-space coupled co-trapped ions \\using telecom photon interference}

\author{Max Bergerhoff}
\affiliation{%
Fachrichtung Physik, Universit\"at des Saarlandes, 66123 Saarbr\"ucken, Germany 
}%
\affiliation{%
Zentrum für Quantentechnologien (QuTe), Universität des Saarlandes, 66123 Saarbr\"ucken, Germany 
}%
\author{Pascal Baumgart}
\affiliation{%
Fachrichtung Physik, Universit\"at des Saarlandes, 66123 Saarbr\"ucken, Germany 
}%
\affiliation{%
Zentrum für Quantentechnologien (QuTe), Universität des Saarlandes, 66123 Saarbr\"ucken, Germany 
}%
\author{Christian Haen}%
\affiliation{%
Fachrichtung Physik, Universit\"at des Saarlandes, 66123 Saarbr\"ucken, Germany 
}%
\affiliation{%
Zentrum für Quantentechnologien (QuTe), Universität des Saarlandes, 66123 Saarbr\"ucken, Germany 
}%
\author{Jonas Meiers}%
\affiliation{%
Fachrichtung Physik, Universit\"at des Saarlandes, 66123 Saarbr\"ucken, Germany 
}%
\affiliation{%
Zentrum für Quantentechnologien (QuTe), Universität des Saarlandes, 66123 Saarbr\"ucken, Germany 
}%
\author{Tobias Bauer}%
\affiliation{%
Fachrichtung Physik, Universit\"at des Saarlandes, 66123 Saarbr\"ucken, Germany 
}%
\affiliation{%
Zentrum für Quantentechnologien (QuTe), Universität des Saarlandes, 66123 Saarbr\"ucken, Germany 
}%
\author{Jonas Haferkamp}%
\thanks{Current address: Fakultät für Informatik, Ruhr-Universität Bochum, 44801 Bochum, Germany}
\affiliation{%
Fachrichtung Mathematik, Universit\"at des Saarlandes, 66123 Saarbr\"ucken, Germany 
}%

\author{Christoph Becher}%
\affiliation{%
Fachrichtung Physik, Universit\"at des Saarlandes, 66123 Saarbr\"ucken, Germany 
}%
\affiliation{%
Zentrum für Quantentechnologien (QuTe), Universität des Saarlandes, 66123 Saarbr\"ucken, Germany 
}%
\author{J\"urgen Eschner}%
\email{juergen.eschner@physik.uni-saarland.de}
\affiliation{%
Fachrichtung Physik, Universit\"at des Saarlandes, 66123 Saarbr\"ucken, Germany 
}%
\affiliation{%
Zentrum für Quantentechnologien (QuTe), Universität des Saarlandes, 66123 Saarbr\"ucken, Germany 
}%

\date{\today}

\begin{abstract}
A quantum repeater segment is a basic building block of a quantum repeater, generating buffered entanglement of quantum memories to connect quantum repeater cells. It also enables the connection between quantum computers. In the implementation we present here, photons emitted from two co-trapped free-space coupled $^{40}$Ca$^+$ ions are converted to the telecom-C band and interfered after transmission over 440\,m of optical fiber (220\,m per arm), where a photonic Bell measurement is performed to create entanglement between the memories.
With this scheme we generate an entangled $\ket{\Psi^+}$ Bell state with $\ge 68(8)$\% fidelity, highlighting trapped $^{40}$Ca$^+$ ions as a promising quantum repeater hardware platform.
\end{abstract}


    \maketitle

\section{Introduction}
In order to realize a quantum network \cite{Kimble_2008} for quantum communication or for distributed quantum computing \cite{DiVincenzo_2000, Nigmatullin_2016, Jiang_2007, Cirac_1999, Monroe_2014}, it is necessary to connect quantum memories, typically atoms or atom-like systems, by flying qubits, typically photons. Due to losses in optical fibers, quantum repeaters \cite{Briegel_1998} are necessary for bridging large distances. Their implementation is pursued with various architectures on the leading qubit platforms such as 
quantum dots \cite{Strobel_2025}, color centers in diamond \cite{Stolk_2024}, neutral atoms \cite{vanLeent_2022}, and trapped ions \cite{Krutyanskiy_2023}.

According to \cite{vanLoock_2020}, two fundamental components of a quantum repeater link have been identified: in a quantum repeater segment (QR segment), two memory qubits are linked and entangled via a photonic connection and Bell measurement; in a quantum repeater cell (QR cell), two proximate and interacting memory qubits, each connected to a photonic link, generate entanglement between the photonic end nodes through atomic Bell state measurement.
The QR cell is the important component in terms of scaling \cite{Luong_2016}, but in order to cover arbitrarily large distances, several QR cells must be connected with the use of QR segments. For both components, a sufficiently long coherence time \cite{Bruzewicz_2019} and a high fidelity \cite{Bock_2018} of the memory-photon entanglement are necessary. For the QR segment, it is also important that the photons are indistinguishable in order to enable photonic entanglement swapping. Furthermore, quantum communication over existing fiber infrastructure requires quantum frequency conversion from atomic wavelengths to a telecom band, and the conversion must maintain both the coherence of the entangled memory-photon state \cite{Bock_2018} and the indistinguishability of photons from two memories \cite{vanLeent_2022}.  

These conditions have all been met in quantum networking experiments with trapped ions or atoms \cite{vanLeent_2022, Bock_2018, Krutyanskiy_2023}, which at the same time show promising quantum processing capabilities \cite{Loeschbauer_2025, Main_2025, Bluvstein_2022, Chiu_2025}. This makes these platforms particularly attractive for quantum repeater implementations. 

Versions of a QR cell have been realized with single atoms in a cavity \cite{Langenfeld_2021}, single ions in a centimeter-scale cavity \cite{Krutyanskiy_2023}, and free-space coupled ions \cite{Bergerhoff_2024}. Realizations of QR segments have been presented that target the connection of two quantum processors \cite{Moehring_2007, Stephenson_2020, OReilly_2024} or clocks \cite{Nichol_2022}, although operating at wavelengths not suitable for efficient frequency conversion \cite{Wright_2018, Hannegan_2021, Yu_2025}. Fewer implementations exist at conversion-compatible wavelengths, based on atomic \cite{Hofmann_2012, Zhang_2022, vanLeent_2022} or ionic \cite{Krutyanskiy2_2023, Cui_2025,Liu_2026} systems. To our knowledge only a single experiment with neutral atoms \cite{vanLeent_2022} in distant traps and a single experiment with trapped ions \cite{Liu_2026} have used conversion of the emitted photons to telecom wavelength.

In this manuscript a proof-of-principle implementation of a QR segment is presented, in which two ions are co-trapped in a Paul trap and their emitted photons are individually collected in free space. A photonic Bell state measurement (BSM) is performed after propagation over 440\,m of optical fiber and conversion to 1550\,nm. In contrast to previous experiments with two ions in the same trap \cite{Stephenson_2020, OReilly_2024, Slodicka_2013}, the photons are not emitted on the excitation transition but in a Raman process, and they are converted with high efficiency \cite{Bock_2018, Arenskoetter_2023}, taking advantage of the particular choice of transition and ion species.
Compared to the experiment with calcium ions using conversion of 393-nm photons in \cite{Liu_2026}, our excitation scheme is inverted, and 854-nm photons are emitted, resulting in an approximately doubled conversion efficiency \cite{Arenskoetter_2023, Bauer_2023}.
\begin{figure*}[!]
    \centering 
    \includegraphics[scale =0.9]{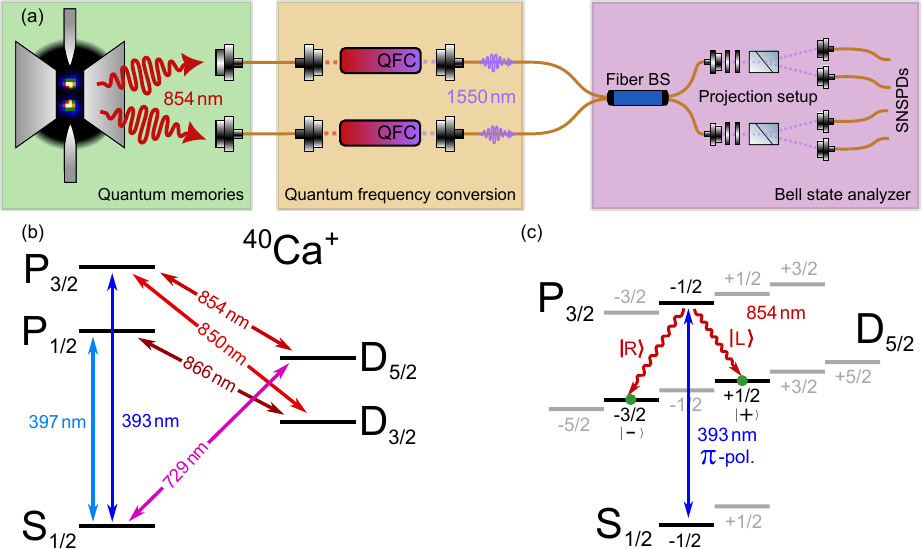}
    \caption{(a) Schematic of the experimental setup, comprising two co-trapped ions in a Paul trap as quantum memories, individual single-photon collection (green), two QFCs (orange), and Bell state analyzer (violet). (b) Level scheme, transitions, and wavelengths of the $^{40}$Ca$^{+}$ ion. (c) Zeeman sub-levels for generation of atom-photon entanglement, showing excitation (blue) and emission paths (red). The relevant levels for generation of the target state (\autoref{eq:APE}) are shown in black. Excitation of the initial state $\ket{\text{S}_{1/2},-1/2}$ to $\ket{\text{P}_{1/2},-1/2}$ with $\pi$-polarized 393-nm light triggers decay to a superposition of $\ket{-}=\ket{\text{D}_{5/2},-3/2}$ and $\ket{+}=\ket{\text{D}_{5/2},+1/2}$ under emission of an 854-nm photon with the corresponding superposition of polarizations 
    $\ket{\text{R}}$ and $\ket{\text{L}}$.} \label{fig:setup}
\end{figure*}
Furthermore, we employ a dual-rail, polarization-encoded scheme, which is robust against phase instabilities; the required polarization preservation over large distance has been demonstrated earlier in \cite{Kucera_2024}. This contrasts with the single-rail entanglement swapping used in \cite{Liu_2026}, which requires phase stability \cite{Zippilli_2008, Dhara_2023}.

Our experiment underscores the suitability of the single-ion platform for future quantum repeater-assisted networks, which utilize atomic quantum memories, efficient telecommunication conversion, and robustness against phase instabilities for scalable quantum communication over long distances. At the same time, this highlights the prospects for interconnecting ion-based quantum processors, using the same set of tools and operations.

In the final discussions, the prospects of using separate ion traps with integrated cavities are explained, as well as the use of the QR segment in a heterogeneous setting, combined with a color center in diamond which brings the need for a common wavelength for interaction and aligned excitation parameters.

\section{Experiment}\label{sec:Experiment}
\subsection{Setup}
In this section we provide a brief description of the experimental setup, shown schematically in \autoref{fig:setup}\,(a); more details are given in our previous works \cite{Bergerhoff_2024,Arenskoetter_2023} and in Appendix \ref{appendix: setup}.

Two $^{40}$Ca$^+$ ions are trapped in a linear Paul trap and Doppler-cooled using laser beams at 397\,nm and 866\,nm (see \autoref{fig:setup}\,(b) for the relevant levels and transitions). After initializing the ions in their ground states with 854-nm and 866-nm laser beams, a train of 15 consecutive laser pulses at 393\,nm is applied to generate single 854-nm photons from both ions. 
Two in-vacuum 0.4-NA high-numerical aperture laser objectives (HALOs) collect the photons along the quantization axis, defined by a magnetic field of $\sim2.85$\,G. On one side of the trap, 397-nm fluorescence photons are coupled into two multi-mode fibers for individual ion state detection, on the other side the photons at 854\,nm are coupled into two single-mode fibers.
The single-mode fibers guide the photons to polarization-preserving quantum frequency converters (QFC) that convert them from 854\,nm to 1550\,nm. The QFC output fibers guide the photons to a Bell state analyzer as shown in \autoref{fig:setup}\,(a), in which the photons are combined on a 50:50 fiber beam splitter (FBS). The FBS outputs are connected to two polarization projection setups which are both set to the R/L basis for entanglement swapping. Photons are registered with superconducting nanowire single-photon detectors (SNSPD) (\textit{Quantum Opus One}) and a time-tagger (\textit{Signadyne, SD-PXE-TDC-H3345-1G}). The distance from each ion to the FBS is around 220\,m.

Quantum state detection at the ions is performed via state-selective fluorescence employing coherent pulses from a 729-nm laser and basis rotations \cite{Bergerhoff_2024,Bock_2018}. For the latter, radio-frequency (RF) pulses produced by a coil in resonance with the transition between the S$_{1/2}$ Zeeman sub-levels are used.

The QFC used for photons of ion 1 is the one described in \cite{Arenskoetter_2023}; for photons of ion 2, a rack-integrated version is used \cite{Bauer_2023}. Their average conversion efficiency is 52\,\%. 
The QFC outputs are each filtered with a Fabry-P\'erot interferometer ($\text{FWHM}=250$\,MHz) and a volume Bragg grating to remove the noise from the conversion. From both QFCs, 6.02\,cts/s of noise are detected in total. Together with the total background of all SNSPDs of 16.22\,cts/s, this results in a signal-to-noise ratio of 13 in the defined acceptance window.
\subsection{Generation of indistinguishable photons}
For successful entanglement swapping, the interfering photons must be indistinguishable. In our system, both ions are co-trapped in a single Paul trap under identical conditions and are excited by the same laser pulse. Consequently, the emitted photons are naturally indistinguishable in frequency. When transitions of the same polarization are selected, they are also indistinguishable in polarization. Temporal indistinguishability is ensured by employing a nanosecond-scale excitation pulse \cite{Baumgart_2026}. After polarization-maintaining quantum frequency conversion, these properties are preserved, as both QFCs are pumped by the same pump laser (in a scenario where the QFCs are separated, a common reference would be needed).

The degree of photon indistinguishability is evaluated using the Hong–Ou–Mandel (HOM) visibility
\begin{align}
    V(T) = 1-\frac{C_\parallel(T)}{C_\perp(T)}.
\end{align}
Here, $C_\parallel(T)$ and $C_\perp(T)$ are the numbers of coincidence events detected at the two outputs of the FBS within a time window of duration $T$, for parallel and orthogonal polarizations, respectively.

\begin{figure}[h]
    \centering 
    \includegraphics[width=\linewidth]{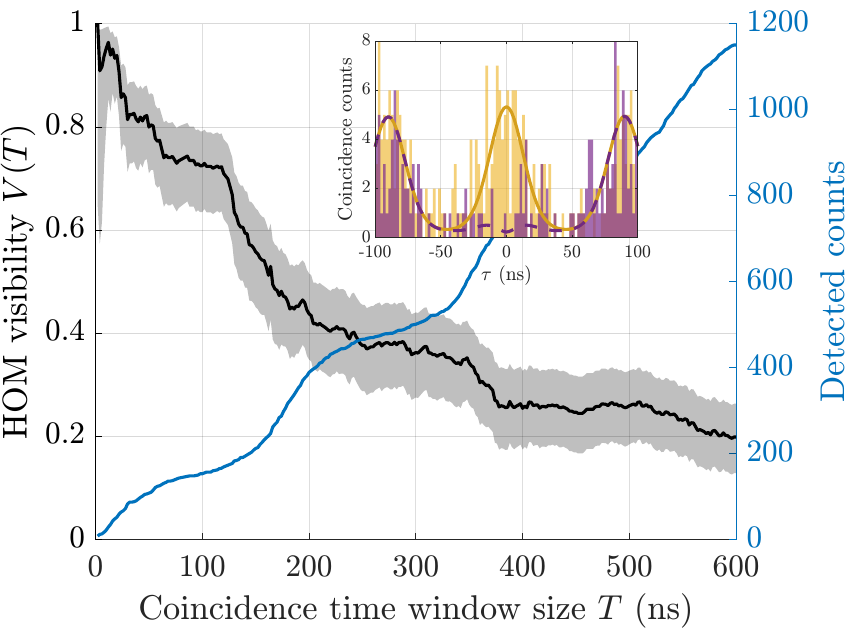}
    \caption{HOM visibility $V(T)$ plotted as a function of the coincidence window size $T$, changed in steps of 1\,ns. Measured data points are shown in black, with their associated uncertainties represented by shaded regions.
    In blue, the number of contributing counts is shown. The inset displays the histograms of coincidence detections as a function of the detection time difference $\tau$ for projections onto orthogonal (yellow bars) and parallel (violet bars) polarizations, evaluated with a bin size of 2\,ns. As a guide to the eye, the lines of the corresponding colors show simulations obtained with the model of \cite{Baumgart_2026} and using the experimentally determined parameters.}  \label{fig:1550HOMV}
\end{figure}

\autoref{fig:1550HOMV} presents the measured HOM visibility as a function of the coincidence window size $T$. Only photonic coincidences followed by successful atomic readout were taken into account (see next section).
The data reveals a decrease in HOM visibility with increasing coincidence window size and a series of plateaus corresponding to the intervals between consecutive excitation pulses. The decrease in HOM visibility after each plateau is attributed to the absence of interference between photons generated by different excitation pulses within the pulse train \cite{Baumgart_2026}. 
Below a time window of 120\,ns, a visibility better than $V(T) = 0.7$ is reached. For shorter time windows, higher visibilities are observed, indicating improved indistinguishability, but at the cost of significantly reduced count rates and thus larger uncertainties. In a trade-off between data rate and visibility, $T=120$\,ns is used for the following verification of entanglement swapping. 

\begin{figure*}[t]
    \centering 
    \includegraphics[scale =0.9]{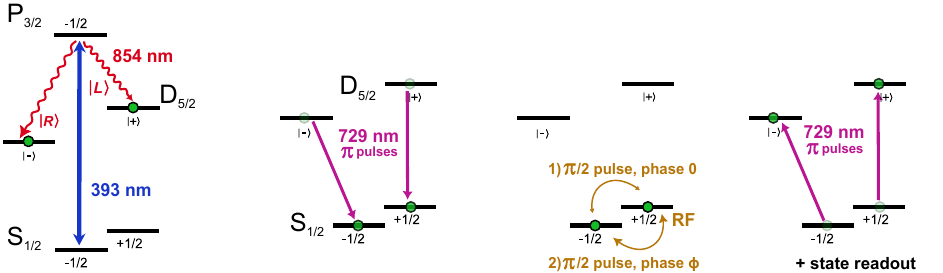}
    \caption{Schematic of the sequence for verification of atom-atom entanglement. Generation of atom-photon entanglement. Population transfer from D$_{5/2}$ to S$_{1/2}$. Basis rotation by the use of two $\pi/2$ RF pulses. Population transfer back from S$_{1/2}$ to D$_{5/2}$ followed by state readout.}  \label{fig:Sequence}
\end{figure*}

\subsection{Atom-photon entanglement}

The entanglement swapping operation of the QR segment relies on initial individual atom-photon entanglement, from which atom-atom entanglement is created by a photonic Bell-state measurement.

Atom-photon entanglement is generated for each ion by simultaneous excitation from the ground state S$_{1/2}$ using 16.6(1)-ns long 393-nm pulses, as already implemented in \cite{Baumgart_2025}.  
An 854-nm photon is emitted during the decay from P$_{1/2}$ to D$_{5/2}$. Through the collection of the photons along the quantization axis and atomic state projection at the end of the sequence, only the final atomic states $\ket{-}=\ket{D_{5/2},-3/2}$ and $\ket{+}=\ket{D_{5/2},+1/2}$ are taken into account. This results in the atom-photon entangled states
\begin{align}
    \ket{\psi}_\text{ap}^j = \sqrt{\frac{2}{3}} \ket{-}_j\ket{\text{R}}_j  + \sqrt{\frac{1}{3}}e^{i (\omega_L t_j+\phi_j)} \ket{+}_j\ket{\text{L}}_j, \label{eq:APE}
\end{align}
with the Larmor frequency $\omega_L = 2\pi \cdot 9.6$\, MHz between the atomic states, the time $t_j$ after photon emission, and the index $j=1,2$ indicating the ion-photon pair. 
The phases $\phi_{1,2}$ arise from differences in the optical path lengths induced by birefringence in the optical components of the setup, predominantly in the optical fibers, also observed in \cite{OReilly_2024}. As this birefringence is compensated by a polarization stabilization routine, the phases are expected to be equalized at the beam splitter (see Appendix \ref{sec:appendix:phase_offset}).

The entanglement between each ion and its corresponding photon is investigated individually by blocking the photon path of the other ion. For a direct measurement of the Clauser–Horne–Shimony–Holt (CHSH) parameter, one polarization projection setup is set to the H/V basis and the other to the D/A basis. To compensate for the unequal Clebsch–Gordan coefficients, losses in $\ket{-}$ are introduced such that the generated state becomes maximally entangled \cite{Bergerhoff_2024}. 
The resulting CHSH parameters, measured at 854\,nm, are 2.42(8) and 2.49(9) for ion~1 and ion~2, respectively. These values are lower than those previously achieved with the same setup using a single ion \cite{Bock_2024}, which is attributed to additional experimental imperfections arising in the two-ion configuration. However, they are consistent with earlier measurements performed with two ions \cite{Baumgart_2025}.

\subsection{Entanglement swapping}

Atom-atom entanglement is marked by coincident detection of two photons of opposite polarization within the acceptance window. The ideal atom-atom state conditioned on such a double detection event is given by
\begin{align}
	\ket{\psi^\pm}_\text{aa} = \frac{1}{\sqrt{2}}\left(\ket{-}_1\ket{+}_2 \pm e^{i(\omega_L\delta t+\phi)}\ket{+}_1\ket{-}_2\right), \label{eq:Psi_aa_phase} 
\end{align}
where the $\pm$-sign depends on whether coincident detection occurred at the same or at opposite outputs of the beam splitter. 
The phase factor contains $\phi=\phi_2-\phi_1$ and an additional phase $\omega_L\delta t$ given by the time difference at the beam splitter 
resulting from the different path lengths that the photons have traveled (effectively a small constant offset of the detection time difference $\tau$). Since $\delta t$ is small in our case (see Appendix \ref{sec:appendix:time_diff}), we use $\omega_L\delta t+\phi \approx 0$.
Hence, the photonic measurement results (imperfections aside) in a projection onto one of the maximally entangled states 
\begin{align}
	\ket{\Psi^{\pm}} = \frac{1}{\sqrt{2}}\left(\ket{-}_1\ket{+}_2 \pm \ket{+}_1\ket{-}_2\right).    
\end{align}
In our setup, a detection of photons in different output ports of the beam splitter leads to a projection on $\ket{\Psi^{+}}$, and a detection in the same output leads to a projection on $\ket{\Psi^{-}}$.

\subsection{Verification}\label{sec:verification}

Atom-atom entanglement is verified through correlations in the two-ion quantum state.
Due to the use of co-trapped ions, operations on them are restricted to global ones. Since only $\ket{\Psi^{+}}$ is not invariant under global rotations, the investigations are done with this state, i.e., after detection of a coincidence event at different output ports of the beam splitter. The protocol for verifying successful entanglement swapping is described in the following and illustrated in \autoref{fig:Sequence}; additional details are given in Appendix \ref{appendix:sequence}.

After coincidence detection, the population from D$_{5/2}$ is transferred to S$_{1/2}$ via two 729-nm $\pi$ pulses. Then a basis rotation is applied by two global RF $\pi/2$ pulses, the first one with a constant phase and the second one with a varying phase $\Phi$. The population is then transferred back to D$_{5/2}$ and, as a last step, state readout is done, distinguishing $\ket{+}$ and $\ket{-}$ via shelving and state-selective fluorescence. 

The optical and RF pulses effectively transform $\ket{\Psi^{+}}$ into $\mathrm{R}(\pi/2,\Phi)\mathrm{R}(\pi/2,0)\ket{\Psi^+}$. Information about entanglement is then obtained from evaluating the parity $P=n_{++}+n_{--}-n_{+-}-n_{-+}$, where $n_{ij}$ are the normalized detection rates of the rotated states, $n_{ij}=N_{ij}/\sum N_{ij}$, calculated from the numbers of detection events $N_{ij}$ with $i,j\in \{+,-\}$ \cite{Slodicka_2013}.

\begin{figure}[h]
    \centering 
    \includegraphics[width=\linewidth]{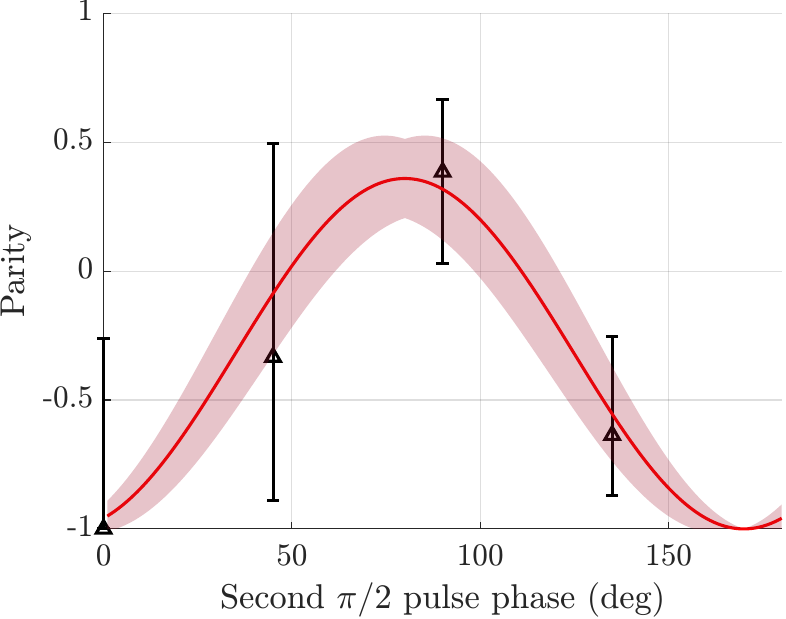}
    \caption{Parity as a function of the phase of second $\pi/2$ pulse. Black points show the measured data at $0^{\circ},45^{\circ},90^{\circ},135^{\circ}$; the red line is a sinusoidal fit with its uncertainty indicated by the red shading.}  \label{fig:1550parity}
\end{figure}

The parity $P$ was measured for four values of the phase of the second RF pulse, $\Phi = 0^{\circ},45^{\circ},90^{\circ},135^{\circ}$, as shown in  \autoref{fig:1550parity}. 
The uncertainties of the individual parity data points result from the underlying binomial distribution. 
From the sinusoidal fit to $P(\Phi)$, a peak-to-peak amplitude of $A=1.36(15)$ is determined. This is clearly larger than the threshold for entanglement of 1.0 \cite{Sackett_2000}. Consequently, entanglement swapping was successful and atom-atom entanglement was established. For comparison, in the case without conversion, a value of 1.56(6) was obtained (see Appendix \ref{appendix:unconverted_measurement}).

From this amplitude, it is possible to calculate a lower bound for the overlap fidelity $F$ of the experimentally prepared state with the ideal $\ket{\Psi^+}$ state, as explained in Appendix \ref{sec:fidelityderivation}. The resulting \autoref{eq:fidelitycomplete} can be written as
\begin{align}
F \ge \frac{1}{2} A. \label{eq:fidelity}
\end{align}
Hence we find a minimal fidelity of $F_{\text{min}} = 68(8)$\,\% for the entanglement swapping using detection of telecom-converted photons, above the threshold of 50\% for entanglement of two particles. The maximal achievable fidelity, derived from our HOM visibility of $V(T) = 0.7$, is given by $F_{\text{max}} = (1+V)/2 = 85\,\%$ \cite{Craddock_2019}.

The uncertainty of the amplitude $A$ and consequently of the fidelity $F$ arises from the individual uncertainties of the data points and thus from the limited amount of data, in total 31 events that were accumulated at a rate of 2.24 $\ket{\Psi^+}$-heralding coincidences per day.

The coincidence detections corresponding to $\ket{\Psi^{-}}$ also herald entangled atom-atom states, even though they may not be used for verification. If these events are included, the overall rate of generating entangled memories increases to 4.7 per day. Both $\ket{\Psi^{+}}$ and $\ket{\Psi^{-}}$ states can be exploited in subsequent quantum networking operations, where knowledge of the specific Bell state might be utilized to incorporate them into further entanglement distribution or swapping protocols.

\section{Summary and Discussion}

In this work, we demonstrate the implementation of a QR segment based on two co-trapped ions. We first characterize the individual atom–photon entanglement via the CHSH parameter and assess photon indistinguishability through the HOM visibility. After the entanglement swapping operation of the QR segment protocol, the resulting atom–atom entanglement is verified by measuring the amplitude of the parity oscillation, giving a value of 1.36(15). This corresponds to a minimum fidelity of 68(8)\,\% of the generated atom-atom state to the $\ket{\Psi^+}$ Bell state. The photonic BSM is performed at 1550\,nm after transmission over 440\,m of optical fiber (220\,m per arm) and quantum frequency conversion of the emitted photons, corroborating the compatibility with future quantum network applications.

The demonstrated telecom-compatible QR segment also represents an important step toward scaling trapped-ion quantum processors through photonic interconnects. Such architectures have been explored at wavelengths less favorable to telecom-conversion \cite{Moehring_2007, Stephenson_2020, OReilly_2024}, but the importance of telecom-wavelength operation is increasingly recognized in the field of trapped-ion quantum computing \cite{Main_2025}. This development is driven by the strong interest in interconnecting multiple quantum processor units via photonic interfaces that can be independently electronically controlled \cite{Loeschbauer_2025}.

The quantum repeater technology realized here also shows a pathway toward a heterogeneous (also referred to as hybrid) QR segment combining a trapped $^{40}$Ca$^+$ ion with a solid-state quantum memory platform, such as a color center in diamond. For long-distance network applications, photons from both memories would be converted to the telecom band. The feasibility of this approach is supported by the present experiment on the ion side and by demonstrations of frequency conversion to telecom wavelength of photons emitted by NV centers \cite{Stolk_2024}, SnV centers \cite{Brevoord_2025, Herrmann_2025prep}, or SiV centers \cite{Knaut_2024, Schaefer_2023}. The main remaining challenges are the compatibility of the photonic qubits and, for some color centers, their entanglement with the emitted photons. Qubit compatibility, and specifically conversion from polarization to time-bin encoding, is already being investigated for the ion platform \cite{Haen_2025prep}.

Future work will aim at realizing a full quantum repeater link \cite{vanLoock_2020} that combines the QR cell \cite{Bergerhoff_2024} and the QR segment, which requires the integration of an additional trapping node and higher-efficiency ion-photon coupling by combining the ion trap with a micrometer-scale optical cavity. Increased generation and detection rates will then enable quantum repeater operation over the Saarbrücken urban fiber testbed \cite{Kucera_2024}.

\section*{Author contributions}
M.\,B. and P.\,B. set up the ion experiment; T.\,B. set up the quantum frequency converter; M.\,B., P.\,B., C.\,H and J.\,M. performed the experiments; M.\,B. and P.\,B. analyzed the data with evaluation scripts written by J.\,M.; J.\,H. derived the inequality for the fidelity; M.\,B. and J.\,E. wrote the paper with input from all authors; C.\,B. supervised the converter part and J.\,E. the ion part of the project.
\begin{acknowledgments}
We gratefully acknowledge financial support from the ''Transformationsprogramm Forschung und Wissenstransfer Saar'' through the Center for Quantum Technologies (QuTe) and from the Federal Ministry of Research, Technology and Space (BMFTR) through projects  Q.sync (16KISQ045), QR.X (16KISQ001K) and QR.N (16KIS2180).
\end{acknowledgments}


\appendix
\section{Details of the experimental setup}\label{appendix: setup}
In this section, different experimental details are described.
\subsection{Detection efficiencies}
The total detection efficiency per excitation attempted of $1.61\cdot10^{-4}$ and $1.58\cdot10^{-4}$ for ion 1 and ion 2 is derivated from an independent measurement.
The contributions of the individual components to detection are listed in \autoref{tab:detection_eff}.

\begin{table}[h]
    \centering
    \setlength{\tabcolsep}{.25cm}
    \caption{Contributions to the detection efficiency of photons emitted by the two ions. The table includes the resulting total efficiencies as well as the independent measured efficiencies.}
    \label{tab:detection_eff}
    \begin{tabular}{l|r|r}
        \toprule
        Contribution to det. efficiency & Ion 1 & Ion 2\\
        \toprule
        $\eta_\text{850}$ & 89.9\,\% & 89.9\,\%\\
        $\eta_\text{mix}$ & 50\,\% & 50\,\%\\
        $\eta_{\sigma}$ & 60\,\%& 60\,\%\\
        $\eta_\text{HALO}$ & 6\,\% & 6\,\% \\
        $\eta_\text{fiber}$ & 19.1\,\% & 22.6\,\% \\
        $\eta_\text{QFC-lab}$ & 25\,\% & 35\,\% \\
        $\eta_\text{filter,FBS}$ & 74\,\% & 52\,\%\\
        $\eta_\text{proj-setup}$ & 78\,\% & 78\,\%\\ 
        $\eta_\text{SNSPDs}$ & 75\,\% & 75\,\%\\
        $\eta_\text{emission}$ & 48\,\% & 48\,\%\\ 
        \botrule
        Total: resulting &  $1.61\cdot10^{-4}$ & $1.87\cdot10^{-4}$ \\
        \botrule
        Total: independent  measured &  $1.61\cdot10^{-4}$ & $1.58\cdot10^{-4}$ \\
    \end{tabular}
\end{table}
The first contributions are calculated using the known atomic properties and those of the objective. The different contributions are as follows: $\eta_{850}$ accounts for parasitic decay from the excited state to D$_{3/2}$ via an 850-nm photon; $\eta_\text{mix}$ arises because only one of the two S$_{1/2}$ Zeeman substates is used for the protocol; and $\eta_{\sigma}$ results from collecting along the quantization axis, thereby suppressing the collection of $\pi$-polarized photons. The HALOs, which have a numerical aperture of 0.4, give $\eta_\text{HALO}$. 

The coupling into the single-mode fibers is determined independently by connecting SNSPDs (\textit{ID Quantique ID281}) for 854\,nm directly to the fibers and measuring the efficiency per excitation attempt, taking into account the previously determined properties and the detection efficiency specified by the manufacturer. This gives $\eta_\text{fiber}$. The transmissions of the different parts of the setup were measured using a laser and resulted in $\eta_\text{QFC-lab}$ for the fiber to the QFC lab, the QFC itself and the fiber back, $\eta_\text{filter,FBS}$ for the filter setup and the FBS, and $\eta_\text{proj-setup}$ for the polarization projection setup. The detection efficiency $\eta_\text{SNSPDs}$ is the average detection efficiency of all four SNSPDs at the optimized bias current, as given by the manufacturer’s datasheet. The probability for photon emission, $\eta_\text{emission}$, is calculated from the photon wave packet over the whole measurement.

These contributions result in a detection efficiency per excitation attempt of $1.61\cdot10^{-4}$ and $1.87\cdot10^{-4}$ for ion 1 and ion 2, respectively, which slightly differs from the values measured independently: $1.61\cdot10^{-4}$ and $1.58\cdot10^{-4}$. This discrepancy can be explained by losses in connectors or instabilities in the coupling during the more than 300\,h of measurement.

\subsection{Complete sequence} \label{appendix:sequence}
In the following the complete sequential implementation of the protocol is listed:
 \begin{itemize}
    \item[(1)] $[ \text{Duration:}\, 3\,\mu\text{s}]$ Doppler cooling with 397\,nm, 866\,nm, in combination with repumping to S$_{1/2}$ with 854\,nm. (next$\rightarrow$(2))
    
    \item[(2)] $[ \text{Duration:}\, 4\,\mu\text{s}]$ 854\,nm photon generation of both atoms with the a train of 15 $16.6(1)\,n\text{s}$ $\pi$-polarized 393\,nm laser pulses. The sequence jumps to (3) conditioned on a detection of a photon emitted by atom\,1, otherwise to (1).

    \item[(3)] $[ \text{Duration:}\, 1.5\,\mu\text{s}]$ Pumping the S$_{1/2}$ population of both atoms to D$_{3/2}$ with the 397\,nm laser. This step, removes the remaining population in the ground state and hides it for the next operations.

    \item[(4)] $[ \text{Duration:}\, 20\,\mu\text{s}]$ Two 729\,nm $\pi$ pulse transfers the $\ket{+}$ and $\ket{-}$ population of both atoms to S$_{1/2}$ $[10\,\mu\text{s}]$.

    \item[(5)] $[ \text{Duration:}\, 10\,\mu\text{s}]$ Two $\pi/2$ pulse RF pulses, the first with constant phase $0^\circ$ and the second with varying phase $\Phi$.

    \item[(6)] $[ \text{Duration:}\, 20\,\mu\text{s}]$ Two 729\,nm $\pi$ pulse transfers S$_{1/2}$ the population of both atoms back to $\ket{+}$ and $\ket{-}$ $[10\,\mu\text{s}]$.

    \item[(7)] $[ \text{Duration:}\, 1.5\,\mu\text{s}]$ Pumping the S$_{1/2}$ population of both atoms to D$_{3/2}$ with the 397\,nm laser. This step, removes the remaining population in the ground state and hides it for the next operations.
    
    \item[(8)] $[ \text{Duration:}\, 70\,\mu\text{s}]$
    Fluorescence detection to detect the population in  D$_{3/2}$. Therefore, the lasers (397\,nm and 866\,nm) are switched on for fluorescence detection and the detected 397\,nm photons are counted. A bright detection indicates the population in D$_{3/2}$ and the sequence is aborted and jumps to (1), otherwise the sequence continues with step (12).

    \item[(9)] $[ \text{Duration:}\, 80\,\mu\text{s}]$ A 729\,nm $\pi$ pulse transfers the $\ket{-}$ population of both atoms to S$_{1/2}$ $[10\,\mu\text{s}]$. A bright result in a following fluorescence detection $[70\,\mu\text{s}]$ projects onto $\ket{-}$ and the sequence is finished, otherwise the sequence continues with step (13).

    \item[(10)] $[ \text{Duration:}\, 80\,\mu\text{s}]$ A 729\,nm $\pi$ pulse transfers the $\ket{+}$ population of both atoms to S$_{1/2}$ $[10\,\mu\text{s}]$. A bright result in a following fluorescence detection $[70\,\mu\text{s}]$ projects onto $\ket{+}$ and the sequence is finished.
 \end{itemize}
 As the detection probability is low, the main contribution to the duration of the sequence comes from the first two steps of 7$\,\mu\text{s}$ which is consistent with the observed duration of a run of 100,000,000 repetitions being around 71.5\,s resulting in a repetition period of 7.15$\,\mu\text{s}$ which gives a rate of around 140,000 s$^{-1}$. The slightly longer experimental repetition period can be attributed to the overhead of the laser power stabilization and the duration of the part of the sequence if a photon was detected.

In order to prepare the entangled memories as a resource for further experiments, the verification sequence would be adapted. Step (4) would selectively address the D$_{5/2}$ Zeeman states D$_{5/2,-1/2}$ and D$_{5/2,+3/2}$, followed by fluorescence detection. Observation of fluorescence from either ion would herald population in the undesired subsystem, originating from initial population in the S$_{1/2,+1/2}$ state, and the protocol would be aborted. Conversely, the absence of fluorescence would indicate the presence of an entangled state within the target subsystem, ready for its utilization in subsequent quantum operations.

\subsection{Polarization control}

Because of around 200\,m of fiber from the ions to the FBS and the polarization detection setup mainly going through non-air-conditioned rooms the polarization must be corrected regularly. The polarization of one ion's photon is corrected using motorized wave plates in one QFC and the polarization of the other ion's photons with motorized wave plates in the polarization projection setup. This is done after 300,000,000 attempts which corresponds to around 36\,min.
Before every correction the R/L ratio were measured using a R polarized input beam to check how good the polarization still was. After the correction this ratio is measured and saved as well.
For the evaluation all of the data acquired between corrections  were taken into account, even if the ratios sometimes dropped from above 1000 to around 200 (sending R polarized light in and measuring the ratios of the detected counts in the R and L output of the polarization projection setup). The HOM visibility in \autoref{fig:1550HOMV} shows that this influence cannot be large.
\subsection{Signal to noise ratio}
The arrival-time distribution of the summed events of all detectors was analyzed. A single-photon wave packet with fifteen peaks was generated by excitation with a train of 15 pulses at 393\,nm, each with a duration of 16.6(1)\,ns. From both QFCs, a total noise count rate of only 6.02\,cts/s is detected. Including the total background from all SNSPDs, amounting to 16.22\,cts/s within the defined acceptance window covering all 15 peaks, this results -- with the given detection probability -- in a measured signal-to-noise ratio of 13.
\subsection{Time difference at FBS}\label{sec:appendix:time_diff}
\begin{figure}[h]
    \centering 
    \includegraphics[width=\linewidth]{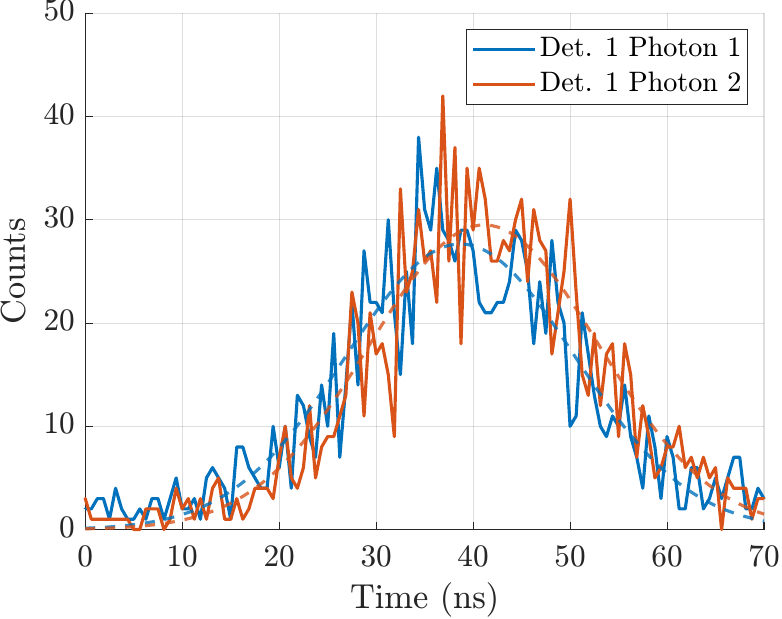}
    \caption{One of the fifteen peaks in the single-photon wave packets from both ions for one selected detector to measure the time difference at the FBS with the fit as dashed line. Photon 1 (2) refers to the photon emitted by Ion 1 (2).}  \label{fig:1550FBStimediff}
\end{figure}
To assess the temporal overlap of the two photon wave packets at the beam splitter, the wave packets are measured sequentially using the same setup, with one input blocked at a time. The fiber lengths from the couplers to the FBS have been equalized as much as possible using 1\,m fibers as smallest steps. In the optimal case, the temporal delay is extracted by comparing the detector signals for both photons. From fits with a gaussian function of the peaks of the wave packets (with the same width), a mean time difference of $\bar{\delta t}=2.5(7)\,\mathrm{ns}$ is obtained, which corresponds -- with the Larmor period of 104.25\,ns -- to a phase difference of 8.6(2.4)$^\circ$. It is therefore justified to assume only a small temporal mismatch. An exemplary comparison for one detector is shown in \autoref{fig:1550FBStimediff}, in which the overlap can already be seen by eye.

The small temporal mismatch is negligible for the HOM visibility as can be seen in \autoref{fig:1550HOMV}, in particular for coincidence time window sizes larger than the lifetime of the excited state, when the temporal extent becomes significant due to back-decays.
\subsection{Phase offset verification}\label{sec:appendix:phase_offset}
\begin{figure}[h]
    \centering 
    \includegraphics[width=\linewidth]{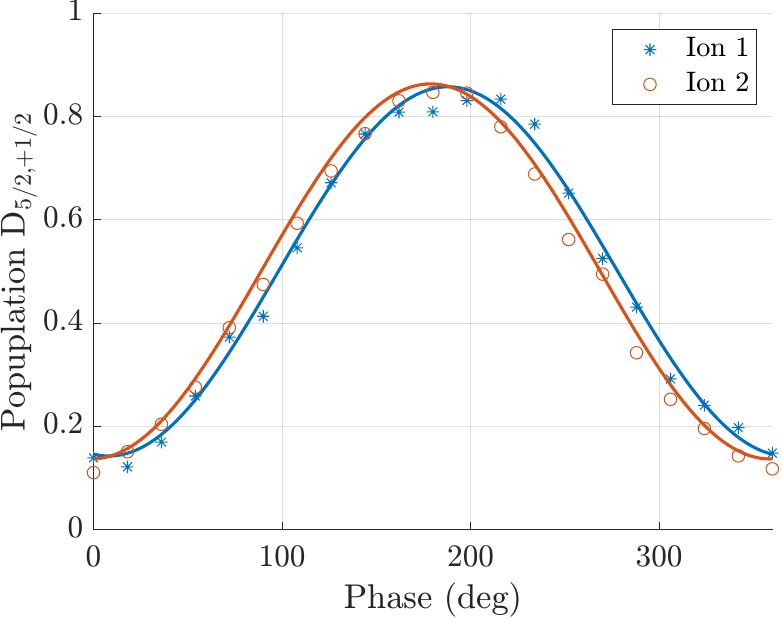}
    \caption{Measurement of ion–photon coherence for ion 1 (blue) and ion 2 (red), conditioned on the detection of V-polarized photons, used to determine the phase difference in the atom–photon entanglement. The solid lines are fits to the sinusoidal functions.}  \label{fig:1550phaseions}
\end{figure}

To verify that the phases in \autoref{eq:APE} are equal, the so-called ion–photon coherence is investigated. For this purpose, ion–photon entanglement is analyzed individually for each ion. The photons are detected in the H/V basis, while the phase of the entangled state is scanned by the Larmor precession. For this purpose, a 3.7-$\mu$s long excitation pulse is used. The results for both ions are shown in \autoref{fig:1550phaseions}.

The oscillations are fitted to a sinusoidal function of the form $0.5 + V \cos(\alpha + \alpha_0)$, where $V$ is the visibility, $\alpha$ is the phase resulting from the Larmor precession, and $\alpha_0$ is the phase offset. From these fits, the difference in the phase offsets is directly extracted. The mean phase difference between ion 1 and ion 2 obtained from the fits is $\phi=-11(1)^\circ$. This discrepancy is most likely attributable to different compensation of polarization rotations in the two photon paths. A dependence of the phase difference on the quality of the polarization correction has already been observed for the case of using unconverted photons, in which the polarization is much more stable, which makes it easier to investigate different qualities of polarization correction. Above all, it is possible to achieve a level of compensation that is not possible in the case of automated stabilization for the scenario with converted photons.

With the result of this section and Appendix \ref{sec:appendix:time_diff} the phase in \autoref{eq:Psi_aa_phase} is obtained as $\delta t\omega_L+\phi =8.6(2.4)^\circ -11(1)^\circ =-2.4(3.4)$ which is consistent with zero within the uncertainty.
\section{Results with unconverted photons}\label{appendix:unconverted_measurement}
As a reference the same measurements without conversion of the photons have been performed.
The difference in the setup with respect to the one used for performing the entanglement swapping in the main part is the absence of the QFCs. Consequently the FBS is one for 854\,nm and the photonic detection in done with different SNSPDs (\textit{ID Quantique ID281}) optimized for 854\,nm.
\begin{figure}[h]
    \centering 
    \includegraphics[width=\linewidth]{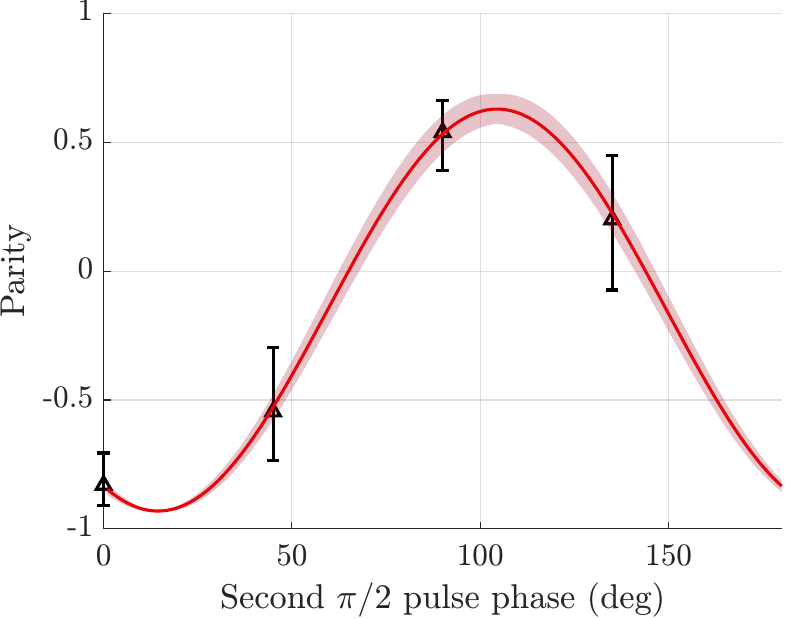}
    \caption{Parity as a function of the phase of second $\pi/2$ pulse for the case of unconverted photons. Black points show the measured data; the red line is a sinusoidal fit with its uncertainty indicated by the red shading.}  \label{fig:854parity}
\end{figure}
This is then the same setup as in \cite{Bergerhoff_2024}, except for the FBS.
As the distances which has to be overcome are in the orders of meters and the fibers are in a single air-conditioned room, the polarization correction is done by hand.

The parity $P$ associated with coincidence detections within a time window of $T=70$\,ns is then measured as a function of the phase of the second RF pulse, as shown in \autoref{fig:854parity}. From the sinusoidal fit an amplitude of 1.56(6) can be determined. This gives with \autoref{eq:fidelity} a fidelity of more then 78(3)\,\%.
\section{Derivation of the fidelity formula} \label{sec:fidelityderivation}
In this section, the formula to calculate the fidelity by measuring the parity after two $\pi/2$ rotations is derived, which is used in the main part as  \autoref{eq:fidelity}.
Here, we consider measurements of the observables
\begin{align}
P(\phi)
&= U(0)U(\phi) \otimes U(0)U(\phi) \notag\\
&\quad \times\, Z \otimes Z \notag\\
&\quad \times\, U^{\dagger}(\phi)U^{\dagger}(0)
\otimes U^{\dagger}(\phi)U^{\dagger}(0),
\label{eq:Pphi}
\end{align}
i.e. the parity operator after local rotations, where $X,Y,Z$ denote the single-qubit Pauli operators and $U(\phi)$ acts on a single qubit, with
\begin{align}
U(\phi)
&= \exp\Big[-i \tfrac{\pi}{4}\big(\cos(\phi) X + \sin(\phi) Y\big)\Big]. 
\label{eq:Uphi}
\end{align}
Ideally, $U(\phi)\otimes U(\phi)$ corresponds to the 
experimentally implemented global operation $R(\pi/2,\phi)$, 
as outlined in \autoref{sec:verification}. 

Our goal will be to show that an expectation value of close to $-1$ for $\phi=0$ and of $+1$ for $\phi=\pi/2$ implies closeness to the state
\begin{align}
\ket{\Psi^{+}}
&= \frac{1}{\sqrt{2}} \left( \ket{01} + \ket{10} \right).
\label{eq:PsiPlus}
\end{align}
Notice that
\begin{align}
\ket{\Psi^{+}}
&= (\mathbb{I} \otimes X)\ket{\Omega},
\notag\\
\ket{\Omega}
&= \frac{1}{\sqrt{2}}\left( \ket{00} + \ket{11} \right).
\label{eq:Omega}
\end{align}
We use the standard notation for the basis states $\ket{0}$ and $\ket{1}$.

We evaluate the expectation values at $\phi=0$ and $\phi=\pi/2$. First, notice that $U^2(0)= e^{-i\pi/2\,X}= iX$. Therefore,
\begin{align}
P(0)
&= Z \otimes Z.
\label{eq:P0}
\end{align}

We have $Z e^{-iaX} = e^{iaX} Z$ for any $a\in\mathbb{C}$ and similarly for other pairs of Pauli matrices. Thus,
\begin{align}
P(\pi/2)
&= (Z \otimes Z) \notag\\
&\quad \times \left(e^{i\pi/4 X} \otimes e^{i\pi/4 X}\right) \notag\\
&\quad \times \left(iY \otimes iY\right) \notag\\
&\quad \times \left(e^{i\pi/4 X} \otimes e^{i\pi/4 X}\right)
\label{eq:Ppi2a}
\\
&= (-i)ZY \otimes (-i)ZY
\label{eq:Ppi2b}
\\
&= X \otimes X.
\label{eq:Ppi2}
\end{align}

We define the observable
\begin{align}
T
&= \frac{1}{2}\left( X \otimes X - Z \otimes Z \right),
\label{eq:Tobs}
\end{align}
which has eigenvalues $+1,0,0,-1$ each corresponding to one of the maximally entangled states
$(\mathbb{I}\otimes  Z^a X^b)\ket{\Omega}$ for $a,b \in \{0,1\}$. We can easily check that the $+1$ eigenvalue corresponds to $\ket{\Psi^{+}}$. Consequently, $T$ satisfies the operator inequality
\begin{align}
T
&\le \ket{\Psi^{+}}\!\bra{\Psi^{+}}.
\label{eq:ineq}
\end{align}

From this observation we can infer bounds on the fidelity of a state $\rho$ via the expectation values $\langle P(\phi)\rangle_{\rho}$. Then,
\begin{align}
F
&= \mathrm{Tr}\big[\rho\,\ket{\Psi^{+}}\!\bra{\Psi^{+}}\big] \notag\\
\notag&\ge \mathrm{Tr}[\rho\, T]
\notag\\
&= \frac{1}{2}\left(
\langle P(\pi/2)\rangle_{\rho}
- \langle P(0)\rangle_{\rho}
\right).
\label{eq:fidelitycomplete}
\end{align}
For the practical use of this formula $P(\pi/2)\rangle_{\rho} - \langle P(0)\rangle_{\rho}$ can be identified as the amplitude $A$ of the parity oscillation as it is introduced in \autoref{sec:verification}.
\bibliography{Bibliography}

\end{document}